\documentclass[
reprint,
nofootinbib,
aps, pra
]{revtex4-1}
\usepackage{custom}
\usepackage{comment}
\usepackage{float}
\usepackage{graphicx} 
\usepackage{multirow}
\usepackage[caption=false]{subfig}

\usepackage{float}
\makeatletter
\let\newfloat\newfloat@ltx
\makeatother

\usepackage{algorithm}
\usepackage[noend]{algpseudocode}

\begin{document}

\title{Truncation technique for variational quantum eigensolver for Molecular Hamiltonians}

\author{Qidong Xu}
\author{Kanav Setia$^1$}
\email{kanavsetia@qbraid.com}
\affiliation{$^1$qBraid Co., 111 S Wacker Dr., Chicago, IL 60606, USA}

\date{\today}

\begin{abstract}
The variational quantum eigensolver (VQE) is one of the most promising quantum algorithms for the near-term noisy intermediate-scale
quantum (NISQ) devices. The VQE typically involves finding minimum energy of a quantum Hamiltonian through classical optimization of a parametrized quantum ansatz. One of the bottlenecks in VQEs is the number of quantum circuits to be measured. In this work, we propose a physically intuitive truncation technique that starts the optimization procedure with a truncated Hamiltonian and then gradually transitions to the optimization for the original Hamiltonian via an operator classification method. This strategy allows us to reduce the required number of evaluations for the expectation value of Hamiltonian on a quantum computer. The reduction in required quantum resources for our strategy is substantial and likely scales with the system size. With numerical simulations, we demonstrate our method for various molecular systems.

\end{abstract}

\maketitle

\section{Introduction}
Recent years have seen growing interest in variational quantum algorithms (VQA) \cite{peruzzo2014variational}  for applications in diverse fields \cite{Cerezo_2021,  Amaro_2022, BravoPrieto2023variationalquantum, Blunt_2022, Landman2022quantummethods}, as they are particularly suitable to study the potential advantage of quantum computation in the current noisy intermediate-scale quantum (NISQ) computers stage. VQAs are typically used to calculate the ground state energy of a Hamiltonian dictated by the problem of interest.  For such problems, it is often efficient to calculate the expectation value of a given observable of the Hamiltonian for a parameterized quantum state on a quantum computer. The expectation value of the Hamiltonian is then used to calculate the updated parameters with the help of an optimizer on a classical computer. The evaluation of the expectation value of the Hamiltonian, followed by the calculation of updated parameters for the ansatz on a classical computer, forms a loop that is run until this procedure converges to a value.  

Variational Quantum Eigensolver, a subclass of VQAs has been a subject of extensive exploration for quantum chemistry \cite{mcclean2016theory,kandala2017hardware, cao2019quantum,bauer2020quantum,lee2021even}, condensed matter \cite{bravo2020scaling,uvarov2020variational}, and many other areas \cite{cao2018potential,banuls2020simulating,liu2022layer}. This is primarily because there are many known transformations from fermionic and bosonic operators to the qubit operators \cite{jordan1993paulische, somma2005quantum}, and many physics-inspired ansatzes could be constructed easily on a quantum computer \cite{OMalley2016,Wecker2015}. Further, many physics-inspired ansatzes could be efficiently prepared on a quantum computer. 
For recent reviews on VQE and best practices, see \cite{cerezo2021variational,tilly2022variational} and references therein. 


Since the proposal of VQE \cite{peruzzo2014variational}, there have been many extensive studies investigating the best ansatz \cite{grimsley2019adaptive, Mehendale_2023}, the classical optimizers used \cite{arrasmith_operator_2020, lavrijsen2020classical}, and the convergence \cite{Cerezo_2021_barren_plateaus,Arrasmith2021effectofbarren}  for VQE. The choice of the ansatz to build parameterized quantum circuits is of particular importance as it typically can only span a subspace of the full Hilbert space and, therefore, essentially sets the upper limit for the distance between the final output minimal value found by the optimizer to the true minimal cost function values corresponding to the ground energy of the Hamiltonian. It also impacts the convergence speed for optimization. Deeper (more complicated) quantum circuits allow for more precise final results as they can explore larger subspaces, but the fact that we are still in the NISQ age with noisy qubits also means that deeper circuits are more vulnerable to noise-induced errors. There have been many recent exploration in this area, trying to achieve the optimal balance between these two issues \cite{lee2018generalized,ryabinkin2018qubit,grimsley2019adaptive,lang2020unitary,ryabinkin2020iterative,tang2021qubit,huang2022robust}. For the classical optimization component, studies have been focusing on how to improve the performance of the classical optimizer, including achieving faster convergence and robustness against the sampling and gate noise in quantum calculation \cite{spall1992multivariate,kandala2017hardware,nakanishi2020sequential,sung2020using,koczor2022quantum,Kubler2020adaptiveoptimizer,sim2021adaptive}. Since there already exists many optimizers in the long study of optimization field, several works have also been devoted to comparing the performance of these classical optimizers\cite{nannicini2019performance,lavrijsen2020classical,pellow2021comparison,bonet2023performance}.

One area where there has been some work but more effort is needed is the number of terms to be measured for the Hamiltonian. Previous works in this area have focused on strategies where the number of circuits to be run is constrained. In this work, we propose a new strategy for reducing the number of measurements for the circuits to be run in the optimization routine in VQE. Our strategy does not depend on the optimization routine and is compatible with many optimizers. It recognizes that many quantum chemistry Fermionic Hamiltonians, when transformed to qubit operators, result in a large number of qubit operator terms with small coefficients. This allows us to engineer an adaptive strategy, wherein we start with a truncated Hamiltonian, which contains a much smaller number of terms, and gradually increase the number of terms in the qubit operator Hamiltonian during the optimization procedure. In this work, we propose two methods for obtaining the truncated Hamiltonian and discuss strategies for transitioning from truncated Hamiltonian to full Hamiltonian to reduce the number of terms to be measured during the optimization procedure. The first one depends on simple cutoffs which serves as a motivational illustration purpose, and in the second method, we propose a systematic truncation strategy via an operator classification with clear physics intuition.

This paper is organized as follows. In Sec.\ref{sec2}, we lay out the necessary backgrounds for the VQE. In Sec.\ref{sec3}, we describe two algorithms based on the idea of Hamiltonian truncation for VQE. We start with a naive algorithm for illustrative and motivational purposes, and we discuss its effectiveness with numerical results validating a simple case of the algorithm. We then analyze the shortcomings of our naive algorithm and present the main result in this work: a truncation algorithm based on the operator classification method, where we also provide numerical simulation results to demonstrate its performance in comparison with the standard VQE algorithm. We conclude in section \ref{sec:conlustion} and discuss some future directions.

\section{Background}
\label{sec2}
We consider a Fermionic Hamiltonian in the second quantization formalism for a given system:
\begin{align}
H_f=\sum_{ij}^M h_{ij}a_{i}^{\dagger}a_{j}+\frac12 \sum_{ijkl}^M h_{ijkl}a_{i}^{\dagger}a_{j}^{\dagger}a_{k}a_{l}\label{hyd_ham}
\end{align}
where $M$ is the number of fermionic modes in the system, $h_{ij}$ and $h_{ijkl}$ are one-electron and two-electron integrals of the operators projected into a given basis set, and $a_i^{\dagger}$, and $a_i$ are fermionic creation and annihilation operators satisfying the following anti-commutation relations:
\begin{align}
\left\{a_i, a_j^\dag\right\}=\delta_{ij}\mathbf{1}\label{ac_relations}.
\end{align}
One of the main problems of relevance in quantum chemistry is the calculation of the energy of the ground-state and excited states of the above Hamiltonian in Eq. (\ref{hyd_ham}). Finding these energies on classical computers is generally intractable because the Fock space associated with the Hamiltonian above scales exponentially, requiring exponential computational resources to store the general state of fermions and calculate the energies of various quantum states. Whereas on a quantum computer, the number of qubits required to represent the fermionic state scale linearly. 

The Fermionic Hamiltonian, Eq. (\ref{hyd_ham}), could be transformed into a qubit operator Hamiltonian using various fermion-to-qubit transformations in the literature, such as Jordan-Wigner \cite{jordan1993paulische}, Bravyi-Kitaev \cite{Bravyi_2002}, Generalized Superfast Encoding \cite{setia2019}, etc. The transformed Hamiltonian is a weighted sum of qubit operators, which are strings of Paulis, given as follows:

\begin{align}
H_q  = \sum_{k=1}^{K} c_k \hat{P}_k, \label{eq:qubit_ham}
\end{align}
where $c_k$ is the coefficient of the Pauli string $P_k$, and the Pauli string is the tensor product of $N$ Pauli matrices that act on $N$ number of qubits separately:
\begin{align*}
\hat{P}_k = \hat{p}_{1k} \otimes \hat{p}_{2k}...\otimes\hat{p}_{Nk}, 
\end{align*}
where, $\hat{p}_{ik} \in \{I, \sigma_x, \sigma_y, \sigma_z \}$ with $I$ the identity matrix, and $\sigma_x,\sigma_y$, and $\sigma_z$ Pauli matrices. The number of Pauli strings $K$ and the number of qubits $N$ depend on the dimensionality of the original $H_f$ and the fermion-to-qubit mapping used. With the fermionic Hamiltonian $H_f$ transformed to qubit operator Hamiltonian $H_q$, the VQE problem reduces to finding the ground state energy of the qubits Hamiltonian $H_q$. 
In a variational quantum eigensolver, one prepares a parameterized ansatz quantum state that can be constructed using various quantum gates. Starting from the initial state of all qubits being in the $|0\rangle$ ground state, the parameterized quantum circuit can be constructed by applying a unitary $U(\boldsymbol{\theta})$ to the initial state. The expectation value of the Hamiltonian in Eq. (\ref{eq:qubit_ham}) with respect to the prepared parameterized quantum state is obtained by measuring the expectation value of each Pauli string and taking their weighted sum:
\begin{align}
\langle 0|U^\dag(\boldsymbol{\theta}) H_q U(\boldsymbol{\theta}) |0\rangle =\sum_{k=1}^{K}c_k\langle 0| U^\dag(\boldsymbol{\theta}) \hat{P}_k U(\boldsymbol{\theta}) |0\rangle.
\end{align}

Upon the measurement of the expectation value of the Hamiltonian, the parameters are updated by an optimization routine which is run until it converges to a minimum value.
\begin{align}
\boldsymbol{\theta}^* = \min_{\boldsymbol{\theta}} \sum_{k=1}^{K} c_k \langle 0| U^\dag(\boldsymbol{\theta})  \hat{P}_kU(\boldsymbol{\theta}) |0\rangle.
\label{VQEeq}
\end{align}

The optimization algorithm is run on a classical computer. We note that since the parametrized state usually can only span a subspace of the Hilbert space for the qubits, the minimized expectation value can only be regarded as an upper bound for the ground energy; finding an efficient ansatz for the parametrized state that can generate final result close to the true eigenvalue with fast convergence speed is of key importance when designing the VQE experiment. 

\section{Truncation technique for qubits Hamiltonian}
\label{sec3}

As the size of the system grows, one of the thing that is expected to slow the optimization procedure is the number of evaluations (measurements) on a quantum computer. Note that, from Eq. (\ref{hyd_ham}), it is apparent that the number of terms grows as $O(M^4)$. This implies that the cost function, defined in Eq. ~(\ref{VQEeq}), also contains $O(M^4)$ Pauli strings. In recent years, many studies have aimed to reduce the depth of the ansatz and reduce the number of measurements required for each iteration of the VQE \cite{arrasmith_operator_2020}. Further, there have also been studies that have looked at the performance of the different classical optimizers for the cost function defined in Eq.~(\ref{VQEeq}) \cite{nakanishi2020sequential, sung2020using,koczor2022quantum, Kubler2020adaptiveoptimizer, sim2021adaptive}. In this work, we present techniques that complement the previous techniques. 

\begin{figure}
    \centering
    \includegraphics[width=0.8\linewidth]{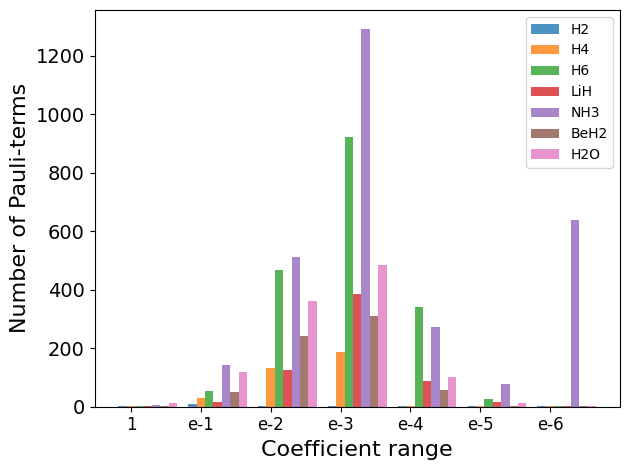}
    \caption{Coefficient range vs. the number of terms for molecular Hamiltonians in STO-3G basis at equilibrium geometry.}
    \label{fig:pauli_term_distri}
\end{figure}

\begin{figure}
    \centering
    \includegraphics[width=0.8\linewidth]{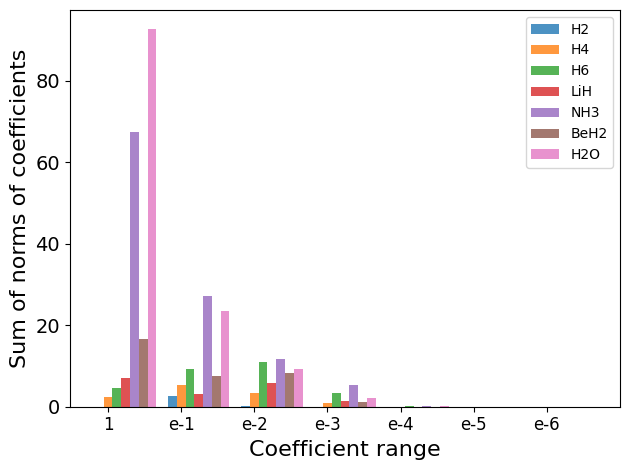}
    \caption{Coefficient range vs. the sum of coefficients within the coefficients range for molecular Hamiltonians in STO-3G basis at equilibrium geometry.}
    \label{fig:coeff_sum}
\end{figure}

Our strategy is based on the observation that even though the cost function usually contains a large number of Pauli strings, many of the coefficients are quite small. Fig. \ref{fig:pauli_term_distri} shows the plot of coefficient ranges with the number of terms in the Hamiltonian. Further, if we sum the norms of the Pauli terms within the coefficients range, it can be seen in Fig. \ref{fig:coeff_sum} that even though the largest terms were very few, they still are dominant, and in fact, the plot for the sum of norms of the coefficients is skewed opposite to the distribution of the number of Pauli-terms. The plot presents the data for small molecules, but this trend gets stronger for larger molecules. This observation suggests that for the Hamiltonians plotted in Fig. \ref{fig:pauli_term_distri}, the terms with coefficients greater than $0.1$ likely contribute the most towards the ground state energy of the system. So, in our algorithm, we start with a qubit Hamiltonian that contains Pauli terms with coefficients above a threshold and start the optimization routine. As the routine progresses, we gradually add more terms with smaller coefficients. Algorithm \ref{alg:basic-truncation} describes our algorithm formally.

\begin{center}
\begin{algorithm}[h]
\caption{Basic Truncation algorithm}\label{alg:basic-truncation}

\begin{algorithmic}[1]

\Function{trunc\_ham}{$\mathcal{H}_q$, cut-off}
\State $\mathcal{H}_{list}\gets \{ \mathcal{H}_q \}$
\For{co in cut-off}
\State $\mathcal{H}_i$=\{$c_kP_k$\} s.t. ($\abs{c_k} P_k \in \mathcal{H}$ and $c_k\geq co$)
\State Append $\mathcal{H}_i$ to $\mathcal{H}_{list}$
\EndFor
\State Reverse $\mathcal{H}_{list}$
\State \Return $\mathcal{H}_{list}$
\EndFunction

\Function{Find-Min}{$\mathcal{H}$, $\ket{\Psi(\hat{\theta})}$}

\State cut-off $\gets$ \{$c_{off1}, c_{off2},..c_{off(k)}$\}
\State $\mathcal{H}_{list} \gets$ $\{\mathcal{H}_{k},..\mathcal{H}_{1},\mathcal{H}_{q}\}$ = trunc\_ham($\mathcal{H}$, cut-off)

\State $\hat{\theta}_0 \gets$ random($len(\hat{\theta})$)

\For{$\mathcal{H}_i$ in $\mathcal{H}_{list}$}
\State E, $\hat{\theta_{i}}$ = VQE($\mathcal{H}_i$,$\ket{\psi(\hat{\theta}_{i-1})}$)

\If{$\mathcal{H}_i == \mathcal{H}_q$}
\State $\hat{\theta}_{opti}=\hat{\theta}_i$
\EndIf

\EndFor
\State \Return $\text{E}, \hat{\theta}_i$
\EndFunction
\end{algorithmic}
\end{algorithm}
\end{center}

To formalize the above ideas, assume that we have a qubit Hamiltonian, $H_q$ given in Eq. (\ref{eq:qubit_ham}) for a VQE problem. We can construct a sequence of Hamiltonians of length $k+1$ derived from $H_q$ such that each Hamiltonian $H_i$ is contained within $H_{i-1}$ and $H_1$ is contained within $H_{q}$:
\begin{align}
H_k \subset H_{k-1} ... \subset H_{1}\subset H_{q}
\end{align}
During the optimization stage, we can optimize the sequence of the Hamiltonians one by one, and when transitioning to the next Hamiltonian, we shall use the previously found best parameters as initial parameters for the next round. For such a strategy to work, the initial Hamiltonian to be optimized $H_k$ needs to contain the Pauli strings in $H_q$ whose absolute value of the coefficients are relatively large compared to other Pauli strings in $H_q$ so that the optimization starts out in the correct direction instead of randomly exploring the parameter space. As more terms that are small enough to be considered perturbations are added, eigenvalues and eigenvectors are stable \cite{tao2010random}. This is why the ansatz parameters for the low energy state of the newly perturbed Hamiltonian could be found in the vicinity of the parameters obtained when the optimization routine converges for the truncated Hamiltonian. 

\subsection{Naive two-step hard cutoff strategy}
\label{ss:naive_algo}

In this subsection, we shall consider a naive hard-cutoff truncation technique. The idea is to choose a hard cutoff $c_{\Lambda}$ and only pick Pauli strings in the original Hamiltonian with coefficients whose absolute value is larger than $c_{\Lambda}$ to obtain a truncated Hamiltonian $H_1$. So, we start with $H_q$ given in Eq. (\ref{eq:qubit_ham}) and obtain the truncated Hamiltonian as
\begin{align}
H_1 = \sum_{k=1}^{N} \Theta(|c_k| - c_{\Lambda}) c_k P_k,
\end{align}
where $\Theta$ is the Heaviside step function. 
The VQE algorithm is started with this truncated Hamiltonian and is allowed to run for a certain number of iterations. Then, without terminating the VQE routine, the truncated Hamiltonian is swapped with the full Hamiltonian, and the VQE routine is continued until convergence or until the max iterations. This two-step truncation technique is a subset of Algorithm 1, and it allows us to test the algorithm and understand performance gains. Our numerical experiments show that this naive strategy delivers considerable improvement in measurement count over the standard VQE algorithm while producing similar final results. It is quite general and does not depend on the choice of the optimizer, ansatz, or a chemical system. Further, this strategy can easily be updated to a more complex strategy with multi-step optimization, given in the previous section, where the truncation thresholds, number of iterations in each step, and convergence criteria all act as hyperparameters.  

In our numerical simulations, we tested various molecular systems that required up to 16 qubits. For each system, we used the variational quantum eigensolver (VQE) algorithm, modified according to our strategy, to calculate the ground energy. We used the Simultaneous Perturbation Stochastic Approximation (SPSA) \cite{Spall_spsa} as our optimizer for the VQE algorithm, as many recent studies have shown that it provides stable and fast performance \cite{sung2020using, lavrijsen2020classical}. It should be noted that our strategy is not optimizer-dependent and should work with any of the optimizers. We explored two different ansatzes for our calculations: Two Local ansatzes and the Unitary Coupled Cluster Singles and Doubles (UCCSD) ansatz \cite{Qiskit,kandala2017hardware}. Two-local ansatz is a heuristic ansatz that can take into account the constraints imposed by the quantum hardware but ignores the physics of the problem, thereby making it hardware efficient, whereas the UCCSD ansatz is an ansatz that captures the physics of the system. There have been many recent studies that strike a balance between Two-Local and UCCSD in the sense that the proposed ansatz tries to capture the physics and the constraints of the hardware simultaneously \cite{ryabinkin2018qubit, grimsley2019adaptive}. Since our strategy works for the two extremes, we expect it to be compatible with other ansatz proposed in the literature as well. 

Table \ref{table:NaiveAlgo} provides the summary of our numerical simulations and demonstrates that even with our naive strategy, an improvement of over 50\% is possible, and the improvement $S$ is calculated with the formula:
\begin{align}
    S = \sum_n k_n*i_n/(K*I), \label{eq:improvement}
\end{align}
where $k_n$ and $i_n$ are the number of terms in the truncated Hamiltonian being used and the number of iterations, respectively, and $K$ is the total terms in the original Hamiltonian $H_q$ with $I = \sum_n i_n$ the maximum iterations used for the standard VQE. For each molecule, we ran 100 simulations to compare the results between our strategy and standard VQE. The performance of VQE depends on various aspects such as variational ansatz, choice of optimizer, etc. So, for our study, we focused on how the results from our strategy compare with the standard VQE. We found that our strategy performs in line with results from standard VQE while delivering significant improvement for the measurement counts. The convergence plots and the distribution of the final minimum eigenvalues for the numerical simulation we performed are given in the appendix \ref{sec:app_A}.

\begin{table}
\begin{ruledtabular}
\centering
\begin{tabular}{cccc}
Molecule & $H_1$/$H_q$ terms &   Improvement\\ \hline
$H_2$ & 11/15 &   14\% \\ 
$H_4$ & 31/361&   48\% \\ 
$H_6$ & 55/1819&  51\% \\ 
$BeH_2$ & 53/666&  49\% \\ 
$H_2O$& 130/1086 & 47\% \\ 
$LiH$ & 18/631 &   51\% \\ 
$NH_3$& 149/2941 & 50\% \\ 
\end{tabular}
\end{ruledtabular}
\caption{ For each of the molecular systems, we provide the improvement in the operator measurement count using the naive truncation strategy. The improvement is calculated using Eq. \ref{eq:improvement}. The maximum iteration for each case is 800, and for the naive hard cutoff strategy, the iterations for $H_1$ and $H_q$ are both 400. The second column gives the number of terms for each optimization stage.} 
\label{table:NaiveAlgo}
\end{table}

\subsection{Operator classification truncation strategy}
\label{ss:op_cl_algo}
In the previous subsection, we see that the truncation strategy can save a significant amount of computation resources and lead to faster optimization while achieving the same level of performance when compared to the standard VQE. When adopting the naive hard cutoff strategy, one has to choose a hard cutoff for the coefficients of the Pauli strings (0.1 in the case of the molecular systems discussed in the previous subsection) to generate the truncated Hamiltonian for the first stage of the optimization. However, such a cutoff value is not available a priori; if we choose the cutoff to be too small, we lose the potential benefit of the truncation strategy as we evaluate more Pauli strings in the first stage, while if we choose the cutoff to be too large, the parameters after the first stage might not be good initial parameters for the second stage (this actually can be seen from several subplots in Fig. \ref{fig:naive truncation convergence} in the appendix \ref{sec:app_A} where a sudden jump appears when transitioning to the second stage optimization). As we discussed, one way to choose the cutoff value is to plot the distribution of the coefficients in the original Hamiltonian $H_0$, from which one can make a reasonable guess. On the other hand, it would be more desirable to have a strategic approach to pick the suitable cutoff. In this section, we propose such an approach to generate the truncated Hamiltonian via operator classification with clear physics motivation.

\begin{figure}[h]
\subfloat[]{\includegraphics[height = 2.1 in]{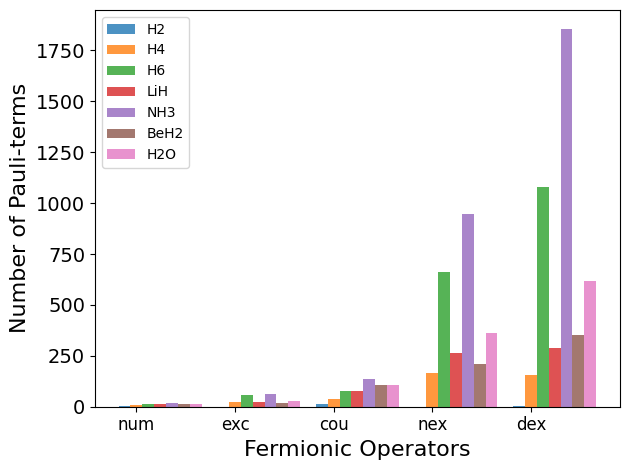}}
\\
\subfloat[]{\includegraphics[height = 2.1 in]{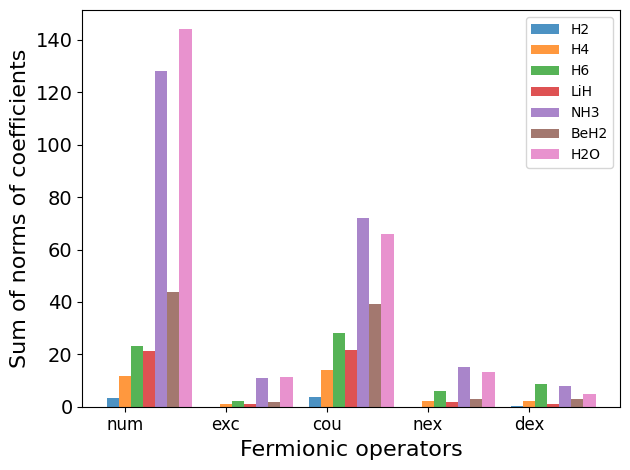}}
\caption{For each of the molecular Hamiltonians in STO-3G basis we plot (a) Number of terms vs each fermionic operator kind  (b) Sum of norm of coefficients vs fermionic operator kind.}
\label{fig:fer_term_distri}
\end{figure}

In Figure \ref{fig:pauli_term_distri}, we analyzed the distribution of the coefficients of Pauli terms. Instead, if we analyze the coefficients of fermionic terms in Eq. (\ref{hyd_ham}), we observe that the biggest contribution comes from the number-operator terms ($a_i^{\dagger}a_i$). This is because the rest of the terms contain an overlap integral, which, depending on the locality of the orbitals in the basis set used, could be relatively small. This motivates segmenting the Hamiltonian into five fermionic operator types: number-operators ($a_{i}^{\dagger}a_i$), excitation-operators ($a_{i}^{\dagger}a_j$), number-excitation operators ($a_{i}^{\dagger}a_k^{\dagger}a_la_i$), coulomb operators ($a_{i}^{\dagger}a_{j}^{\dagger}a_ja_i$), and double-excitation operators ($a_i^{\dagger}a_j^{\dagger}a_ka_l$) \cite{Setia_2018}. It is observed that the greatest contribution to the Hamiltonian comes in the following order (Figure \ref{fig:fer_term_distri}):
\begin{align*}
H_{num}>H_{cou}>H_{exc}> H_{nex}> H_{dex},
\end{align*}
where $H_{num}=\sum^{M}_{i}a_i^{\dagger}a_i$, $H_{cou}=\sum^{M}_{ij}a_i^{\dagger}a_j^{\dagger}a_ja_i$, $H_{exc}=\sum^{M}_{ij}a_i^{\dagger}a_j$, $H_{nex}=\sum^{M}_{ikl}a_i^{\dagger}a_k^{\dagger}a_la_i$, and $H_{dex}=\sum^{M}_{ijkl}a_i^{\dagger}a_j^{\dagger}a_ka_l$

So, instead of defining a truncation threshold and running the optimization, we can make the truncation criteria to be the exclusion of certain fermionic operators from the Hamiltonian. We start the optimization routine with just the number-operator terms, followed by Coulomb operators, excitation operators, number excitation operators, and double-excitation operators. Furthermore, we note that the two leading Hamiltonins in the contribution sequence, the number operator and Coulomb operator terms when transformed to qubit operators using Jordan-Wigner transformation, lead to just $Z-$string qubit operators, indicating all these terms commute and can be measured in a single measurement, we can then start the optimization routine with both of these terms included. The details of the algorithms are given in the Algorithm \ref{alg:phys-insp-truncation}.

This operator-classification approach not only provides a good way of deciding the truncation but also provides some physical meaning to how the Hamiltonian is `evolved' throughout the optimization routine. In the initial phases, the optimization is performed for a Hamiltonian that only includes the potential energy and electronic repulsion terms. Then, in the following steps, the quantum mechanical terms are added. In our testing, we found that this operator classification technique works better than the standard VQE and the truncation technique proposed in the previous section. This is because in first phase of optimization involves only a single measurement. 

\begin{center}
\begin{algorithm}[h]
\caption{Physics inspired truncation algorithm}\label{alg:phys-insp-truncation}
\begin{algorithmic}[1]
\Function{trunc\_ham}{$\mathcal{H}$, cut-off}
\State $\mathcal{H}_{num}$=\{$h_{pq}a_p^{\dagger}a_q$\} s.t. ($h_{pq}a_p^{\dagger}a_q$ $\in$ $\mathcal{H}$, p=q)
\State $\mathcal{H}_{cou}$=\{$h_{pqrs}a_p^{\dagger}a_q^{\dagger}a_ra_s$\} s.t. ($h_{pqrs}a_p^{\dagger}a_q^{\dagger}a_ra_s$  $\in$ $\mathcal{H}$, len(set\{p,q,r,s\})=2)
\State $\mathcal{H}_{exc}$=\{$h_{pq}a_p^{\dagger}a_q$\} s.t. ($h_{pq}a_p^{\dagger}a_q$  $\in$ $\mathcal{H}$, p$\neq$q)
\State $\mathcal{H}_{nex}$=\{$h_{pqrs}a_p^{\dagger}a_q^{\dagger}a_ra_s$\} s.t. ($h_{pqrs}a_p^{\dagger}a_q^{\dagger}a_ra_s$ $\in$ $\mathcal{H}$, len(set\{p,q,r,s\})=3)
\State $\mathcal{H}_{dex}$=\{$h_{pqrs}a_p^{\dagger}a_q^{\dagger}a_ra_s$\} s.t. ($h_{pqrs}a_p^{\dagger}a_q^{\dagger}a_ra_s$ $\in$ $\mathcal{H}$, len(set\{p,q,r,s\})=4)
\State $\mathcal{H}_{list}=\{\mathcal{H}_{num}, \mathcal{H}_{cou},\mathcal{H}_{exc},\mathcal{H}_{nex},\mathcal{H}_{dex}\}$
\State \Return $\mathcal{H}_{list}$
\EndFunction

\Function{Find-Min}{$\mathcal{H}_q$, $\ket{\Psi(\hat{\theta})}$}
\State $\mathcal{H}_{segments} \gets$ $\{\mathcal{H}_{num}, \mathcal{H}_{cou},\mathcal{H}_{exc},\mathcal{H}_{nex},\mathcal{H}_{dex}\}$ = segments($\mathcal{H}_f$)
\State $\mathcal{H}_3 = \mathcal{H}_{num}+ \mathcal{H}_{cou}$, \State$\mathcal{H}_2 = \mathcal{H}_{3}+ \mathcal{H}_{exc}$, \State$\mathcal{H}_1 = \mathcal{H}_{2}+ \mathcal{H}_{nex}$, \State$\mathcal{H}_q = \mathcal{H}_{1}+ \mathcal{H}_{dex}$

\State $\mathcal{H}_{list} = \{\mathcal{H}_3,\mathcal{H}_2,\mathcal{H}_1,\mathcal{H}_q\}$
\State $\hat{\theta}_0 \gets$ random($len(\hat{\theta})$)

\For{$\mathcal{H}_i \in \mathcal{H}_{list}$}
\State E, $\hat{\theta_{i}}$ = VQE(jordan-wigner$(\mathcal{H}_i)$,$\ket{\Psi(\hat{\theta}_{i-1})}$)

\If{$\mathcal{H}_i == \mathcal{H}_f$}
\State $\hat{\theta}_{opti}=\hat{\theta}_i$
\EndIf

\EndFor
\State \Return $\text{E}, \hat{\theta}_i$
\EndFunction
\end{algorithmic}
\end{algorithm}
\end{center}

\begin{table}
\begin{ruledtabular}
\centering
\begin{tabular}{cccc}
Molecule & $H_3/H_2/H_1/H_q$ terms &   Improvement\\ \hline
$H_2$ & 11/ 11/ 11/ 15  &   57\% \\ 
$H_4$ & 37/ 61/ 205/ 361&   68\% \\ 
$H_6$ & 79/ 139/ 739/ 1819&  72\% \\ 
$BeH_2$ & 106/122/314/666, &  70\% \\ 
$H_2O$& 106
/134
/470
/1086 & 71\% \\ 
$LiH$ & 79/ 103/ 343/ 631 &   69\% \\ 
$NH_3$& 137/ 201/ 1085/ 2941 & 73\% \\ 
\end{tabular}
\end{ruledtabular}
\caption{ For each of the molecular systems, we provide the improvement in the operator measurement count using the operator classification truncation strategy. The improvement is calculated using Eq. \ref{eq:improvement}. The maximum iteration for each case is 1000, and the iterations for $H_3$, $H_2$, $H_1$, and $H_q$ are 500, 100, 200 and 200 respectively. The second column gives the number of terms for each optimization stage.}
\label{table:physics-inspired}
\end{table}

For our testing, we tested both the hardware-efficient ansatz and UCCSD. Our proposed strategy worked for both, but we conjecture that our strategy will work particularly best with UCCSD as the system size grows. Table \ref{table:physics-inspired} summarizes our numerical simulations. As is evident from the data, an improvement of over 70\% is achieved for many molecules along with substantial improvement for the rest. From Eq. (\ref{eq:improvement}), notice that the improvement over standard VQE is a function of the number of terms in each step of our new algorithm and the total iterations used for that particular step. In particular, we note that the number of terms in the initial steps is much smaller than the total number of terms that are used in the final step. Therefore, the improvement is also constrained by the number of iterations being used in the final step of the algorithm. As the molecule size increases, we expect the number of iterations required for convergence also to increase, implying that further improvement might be possible.

\section{Conclusion and outlook}
One of the main concerns around the variational quantum eigensolver is the number of quantum circuit measurements required for its implementation. In this work, we proposed two different strategies for reducing the measurement counts and we see both strategies deliver substantial performance gains, indicating a solid step toward addressing the measurement problem in VQE. For future directions, we reckon there is a way to combine both of our strategies for further improvement. The improvement for our physics-inspired truncation strategy is bound by the number of measurements in the last step of the algorithm. In principle, it should be possible to use Algorithm 1 as a subroutine for the last step of Algorithm 2. In this work, we considered the operator classification method for the class of molecular Hamiltonians; it would also be interesting to explore the possibility of applying such kind of strategy to other types of Hamiltonians containing perturbative terms. 

\label{sec:conlustion}

\begin{acknowledgments}
This work is supported by Wellcome Leap as part of the Quantum for Bio Program. 
\end{acknowledgments}

\bibliography{main}
\onecolumngrid

\newpage
\appendix

\section{Convergence plots for naive and the operator classification truncation strategy}
\label{sec:app_A}
In this section, we provide the details of our numerical simulation setup. We explored seven molecular systems, $H_2$, $H_4$, $H_6$, $H_2O$, $LiH$, $H_2O$, $BeH_2$. We used qiskit \cite{Qiskit} along with the pyscf \cite{Sun_2020} driver to run the Hartree-Fock method in STO-3G basis and at equilibrium geometries to obtain the one-body and two-body terms. The geometries are given in the following table:

\begin{table}[h]
    \centering
    \begin{tabular}{c|c}
         Molecular system & Geometry ($\AA$)  \\
         \hline
         $H_2$ & H 0 0 0; H 0 0 0.735 \\
         $H_4$ & H 0 0 0; H 0 0 0.735; H 0 0 1.535; H 0 0 2.135 \\
         $H_6$ & "H 0 0 0; H 0 0 0.735; H 0 0 1.535; H 0 0 2.135; H 0 0 2.835; H 0 0 3.57 \\
         $LiH$ & Li .0 .0 .0; H .0 .0 1.5949 \\
         $NH_3$ & N1 0.0000 0.0000 0.0000;H2 0.0000 -0.9377 -0.3816;H3 0.8121 0.4689 -0.3816;H4 -0.8121 0.4689 -0.3816 \\
         $BeH_2$ & Be1 0.0000 0.0000 0.0000; H2 0.0000 0.0000 1.3264; H3 0.0000 0.0000 -1.3264 \\
         $H_2O$ & O1 0.0000 	0.0000 	0.1173; H2 	0.0000 	0.7572 	-0.4692; H3 	0.0000 	-0.7572 	-0.4692 \\
    \end{tabular}
    \caption{Geometries for various molecular systems. The units for the coordinates are in angstrom.}
    \label{tab:mol_geom}
\end{table}

For all our numerical simulations, we used the Jordan-Wigner transformation to map the fermionic Hamiltonian to the qubit operator Hamiltonian. For simulating the quantum circuits, we use the \texttt{AerEstimator} with statevector simulator available in qiskit. 

\begin{figure}
    \centering
    \subfloat[]{\includegraphics[width=0.33\textwidth]{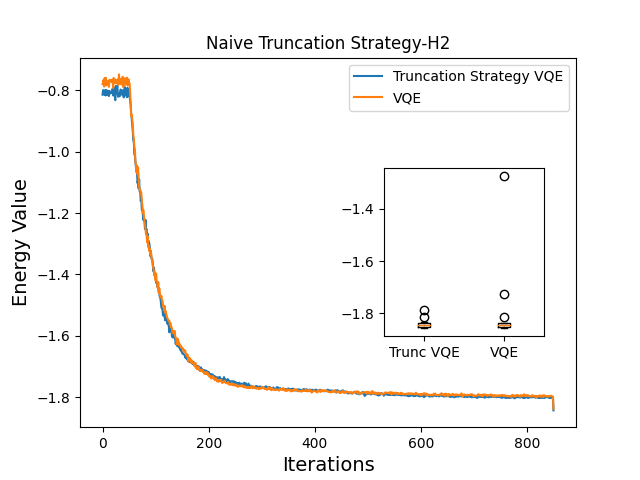}} 
    \subfloat[]{\includegraphics[width=0.33\textwidth]{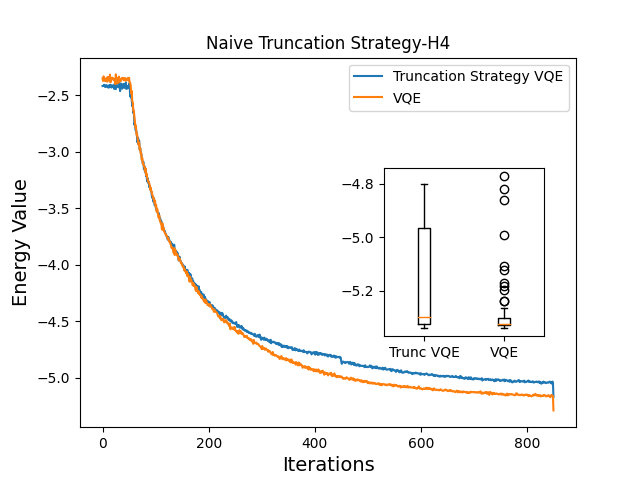}} 
    \subfloat[]{\includegraphics[width=0.33\textwidth]{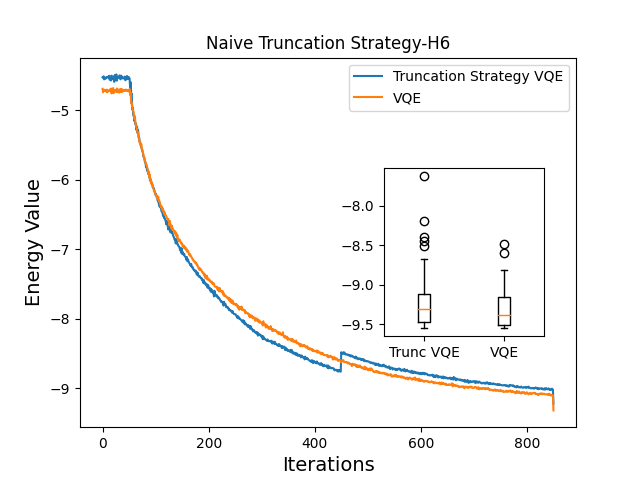}}\\
    \subfloat[]{\includegraphics[width=0.33\textwidth]{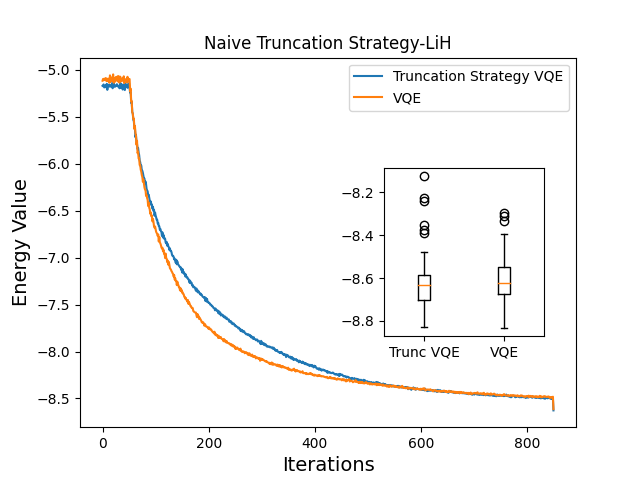}} 
    \subfloat[]{\includegraphics[width=0.33\textwidth]{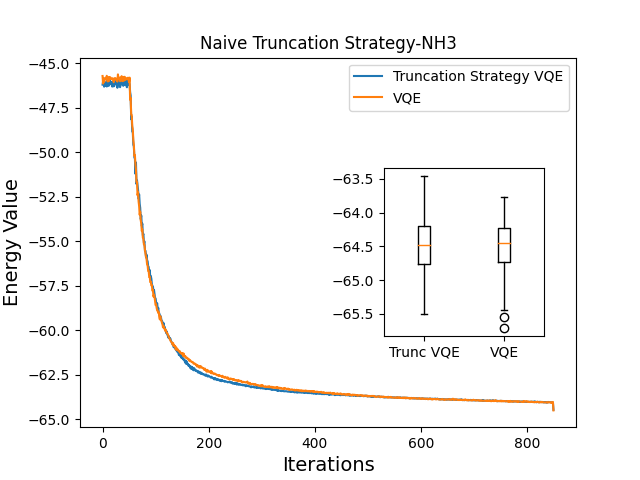}}
    \subfloat[]{\includegraphics[width=0.33\textwidth]{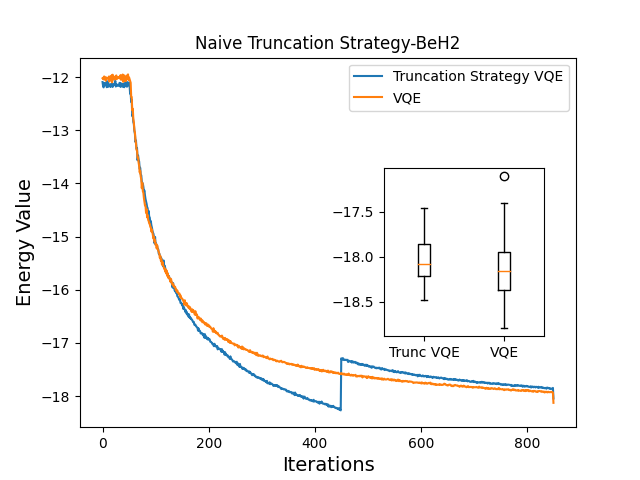}}\\
    \subfloat[]{\includegraphics[width=0.33\textwidth]{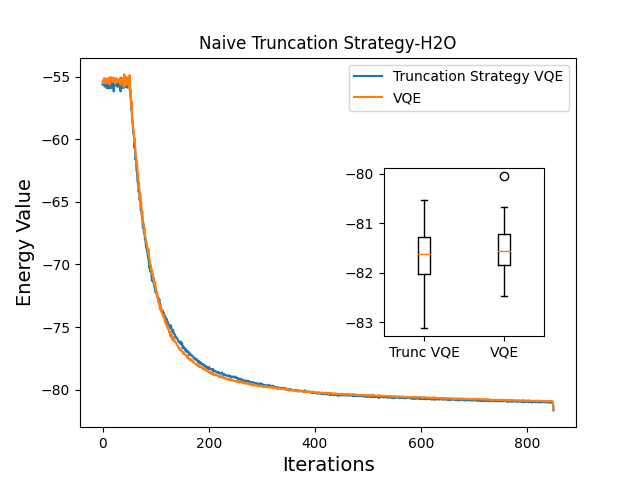}}
    \subfloat[]{\includegraphics[width=0.33\textwidth]{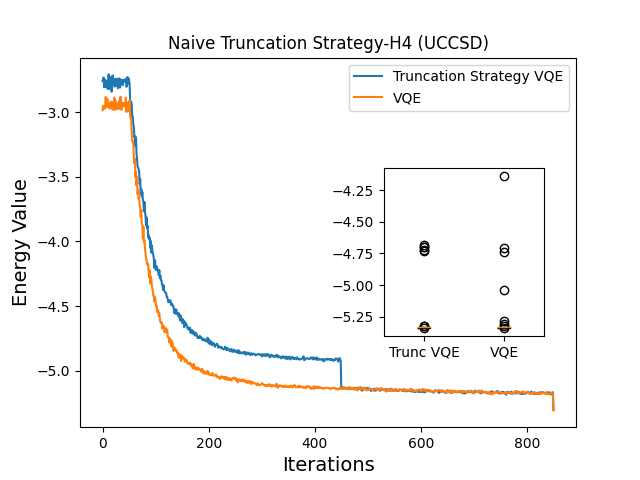}}
    \subfloat[]{\includegraphics[width=0.33\textwidth]{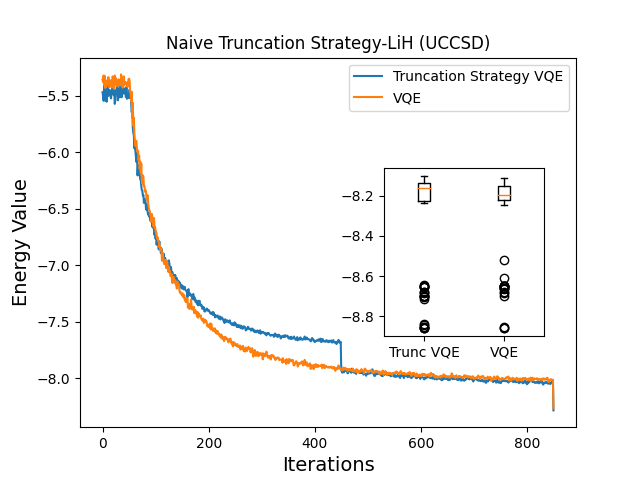}}
    \caption{Convergence comparison plots between the naive two-step strategy and the standard VQE for different molecules. The inset box plots display the distribution for the final minimum eigenvalues. All subplots are obtained by averaging 100 runs of simulations.}
    \label{fig:naive truncation convergence}
\end{figure}

In Fig. \ref{fig:naive truncation convergence} we compare the averaged convergence plots for the naive two-step truncation strategy and the standard VQE for all seven molecules, and we show the comparison of the final minimum eigenvalues found by both strategies in the inset box plot. For both strategies, we ran 100 simulations and took the average for the plots. For simulations in subplots (a) to (g) in Fig. \ref{fig:naive truncation convergence}, we used the \texttt{TwoLocal()} ansatz with $R_y$ rotation blocks and $CZ$ entangling gates, and the number of repetitions is three; for simulations in subplots (h) and (i), we used the UCCSD ansatz.  The classical optimizer used is SPSA, with the maximum iterations set to 800 for the standard VQE and 400 for both Hamiltonians in the naive two-step truncation strategy. We note that the SPSA optimizer has a default initial calibration phase with 50 iterations, we see that there are in total 850 evaluations in the plots. We also note that our convention is slightly different from the hyperparameters `iteration' defined in the SPSA optimizer; one iteration in our convention corresponds to one measurement while one iteration in SPSA corresponds to two evaluations (measurements) of the cost function. From the subplots, we see that the naive two-step method achieves similar convergence speed with a similar distribution of the final minimum eigenvalues when compared with the standard VQE. some kinks can be observed in the subplots because the energy to the subsequent truncated Hamiltonian is higher (or lower), indicating the importance of a proper choice of cutoff value for the original Hamiltonian. 

\begin{figure}
    \centering
    \subfloat[]{\includegraphics[width=0.33\textwidth]{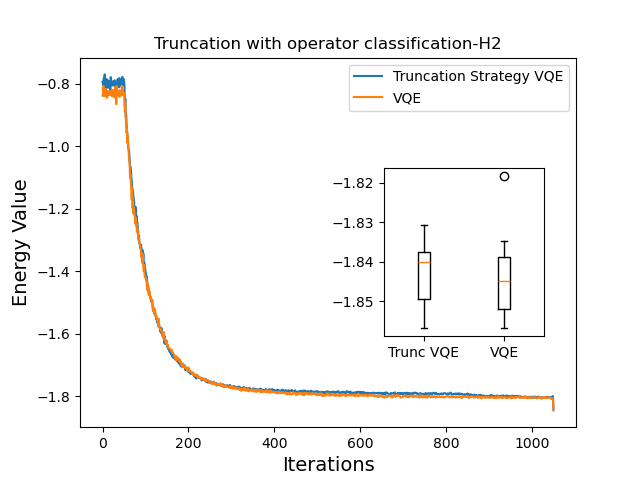}} 
    \subfloat[]{\includegraphics[width=0.33\textwidth]{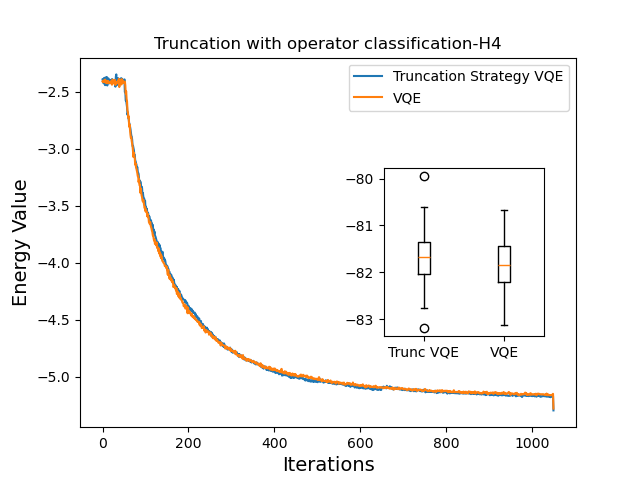}} 
    \subfloat[]{\includegraphics[width=0.33\textwidth]{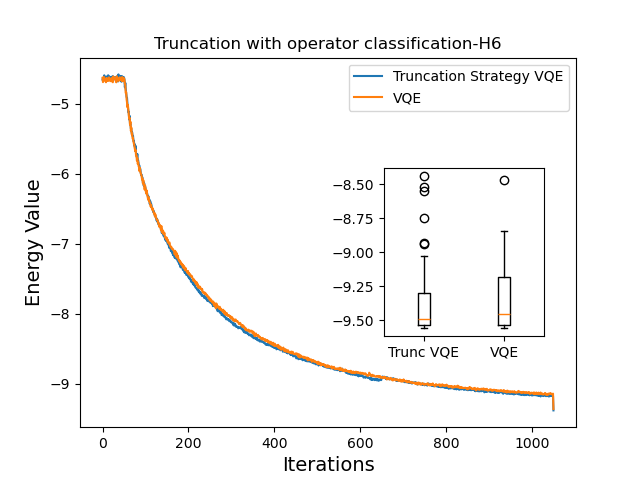}}\\
    \subfloat[]{\includegraphics[width=0.33\textwidth]{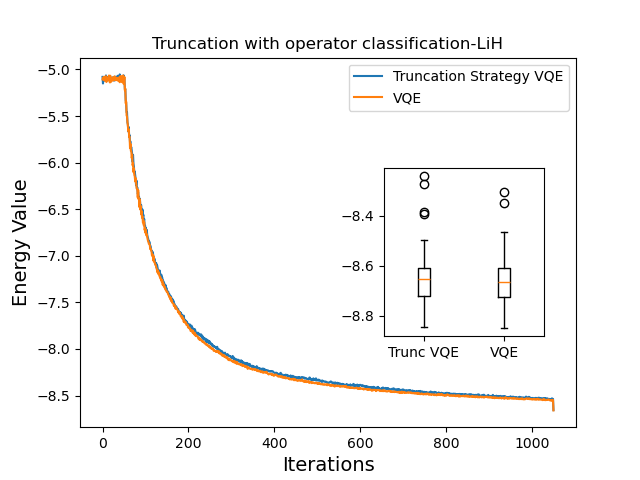}} 
    \subfloat[]{\includegraphics[width=0.33\textwidth]{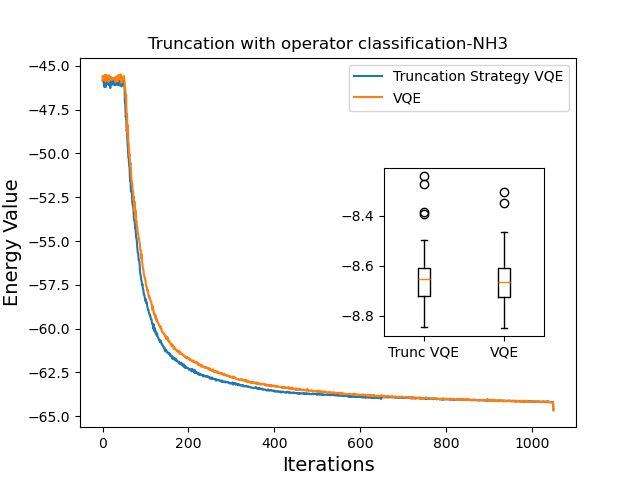}}
    \subfloat[]{\includegraphics[width=0.33\textwidth]{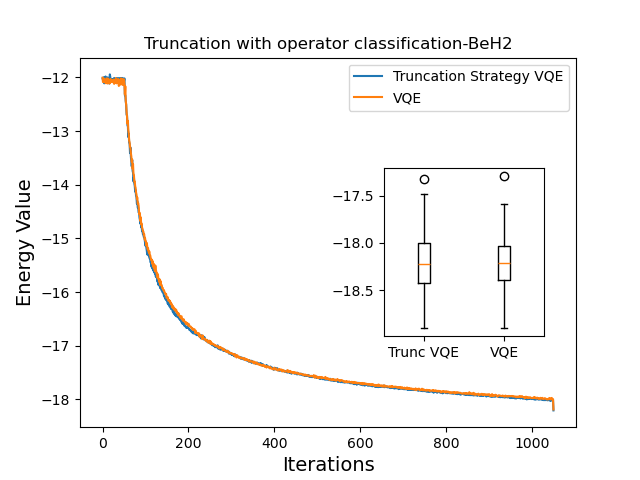}}\\
    \subfloat[]{\includegraphics[width=0.33\textwidth]{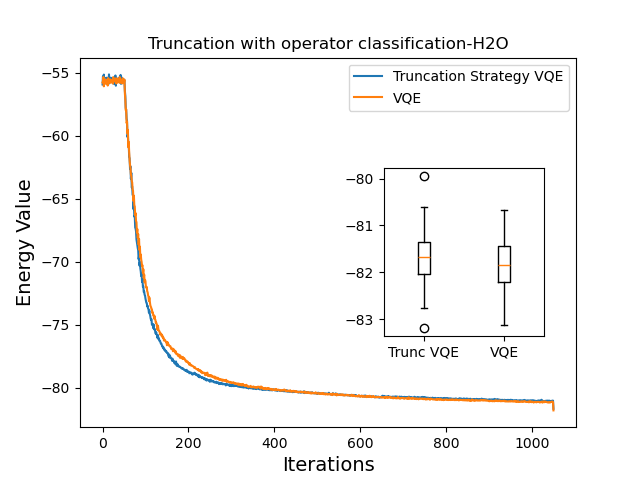}}
    \subfloat[]{\includegraphics[width=0.33\textwidth]{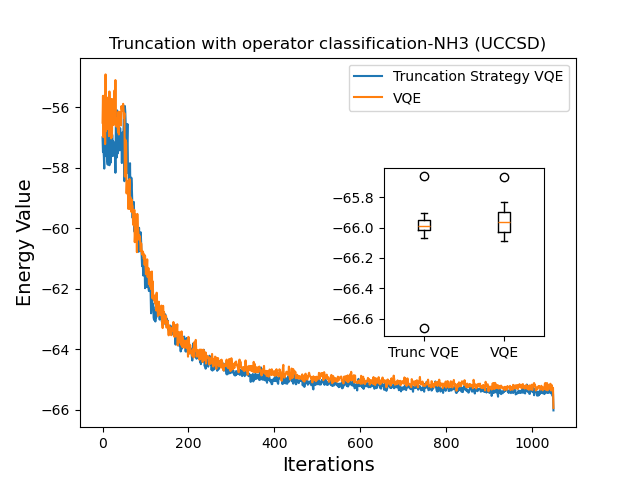}}
    \subfloat[]{\includegraphics[width=0.33\textwidth]{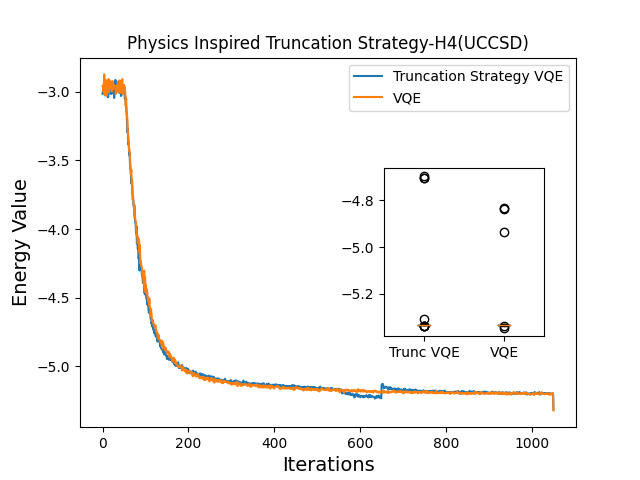}}
    \caption{Convergence comparison plots between the operator classification truncation strategy and the standard VQE for different molecules. The inset box plots display the distribution for the final minimum eigenvalues. All subplots are obtained by averaging 100 runs of simulations except for subplot (h), which is obtained by averaging 10 runs of simulations.}
    \label{fig:op classification truncation convergence}
\end{figure}

In Fig. \ref{fig:op classification truncation convergence} we compare the averaged convergence plots for the operator classification truncation strategy and the standard VQE for all seven molecules with subplots (a) to (g) using the \texttt{TwoLocal()} ansatz with the same settings as in Fig. \ref{fig:naive truncation convergence}, and subplot (h) and (i) using the UCCSD ansatz. All subplots are obtained by averaging 100 simulations, except for subplot (h) which is obtained by averaging 10 simulations due to its much longer clock run time. The classical optimizer used is SPSA, with the maximum iterations set to 1000 for the standard VQE, and 500, 100, 200, and 200 iterations for $H_3$, $H_2$, $H_1$, and $H_q$ respectively. It is of interest to note that the kinks are mostly absent in the operator classification truncation method, indicating that this is a generic algorithm suitable for the study of quantum chemistry VQE problems.

\end{document}